\documentstyle[epsfig]{mn}

\title{Evolution of Brightest Cluster Galaxies in X-Ray Clusters}

\author[Brough et al.]
       {S.~Brough,$^1$\thanks{Email: sb@astro.livjm.ac.uk} C. A.~Collins,$^1$ D. J.~Burke,$^{2,3}$ R. G.~Mann,$^4$ P. D.~Lynam$^5$\\$^1$Astrophysics Research Institute, Liverpool John Moores University, Egerton Wharf, Birkenhead, CH41 1LD, UK\\$^2$Institute for Astronomy, University of Hawaii, 2680 Woodlawn Drive, Honolulu, HI 9682, USA\\$^3$Harvard-Smithsonian Center for Astrophysics, 60 Garden Street, Cambridge, MA 02138, USA\\$^4$Institute for Astronomy, University of Edinburgh, Royal Observatory, Blackford Hill, Edinburgh, EH9 3NJ, UK\\$^5$Max-Planck-Institut f\"{u}r extraterrestrische Physik, Giessenbachstrasse Postfach 1603, D-85741 Garching, Germany}
\date{Accepted... Received...; in original form 2001 September}

\pagerange{\pageref{firstpage}--\pageref{lastpage}}
\pubyear{2001}

\begin{document}

\maketitle

\label{firstpage}

\begin{abstract}

A recent paper (Burke, Collins \& Mann 2000) presents the analysis of the 
K-band Hubble diagram of 76 brightest cluster galaxies (BCGs) in X-ray clusters
 and shows that the properties of BCGs depend on the X-ray luminosity 
($L_{\rm{X}}$) of their host clusters.  Unfortunately, the low numbers of 
nearby clusters in this sample makes it difficult to constrain evolutionary 
trends.  In this letter we extend the Hubble diagram of Burke et al. (2000) 
to a total of 155 clusters using new data on 79 BCGs at $z\leq0.1$ from the 
2MASS extended source catalogue.  We show that the major division between BCGs 
in high and low-$L_{\rm{X}}$ clusters disappears at $z\leq0.1$, with BCGs 
having similar absolute magnitudes independent of the X-ray luminosity of 
their host clusters.  At larger redshifts the K-band light of BCGs in 
high-$L_{\rm{X}}$ systems is consistent with little or no merging back to 
$z\sim0.8$, whereas BCGs in the low-$L_{\rm{X}}$ systems have a different 
evolutionary history, with many increasing their mass by a factor $\geq4$ 
since $z\simeq1$.  This provides direct evidence of hierarchical merging in 
a galaxy population.

\end{abstract}


\section{Introduction}
Brightest cluster galaxies (BCGs) provide a unique sample with which to study 
the evolution of galaxies in a cluster environment.  They are the most 
luminous galaxies emitting purely photospheric light in the universe and are 
uniquely positioned at the centre of the cluster gravitational potential.  
A classic observation of BCGs is that they vary little in luminosity 
within a fixed metric aperture (e.g. Sandage 1972a,b) and in recent years 
the near-infrared K-band Hubble diagram has been analysed to redshifts $z\simeq 1$
(e.g. Arag\'{o}n-Salamanca, Baugh \& Kauffmann 1998; Collins \& Mann 1998). 
The K-band is ideal to study BCG evolution as extinction  
at 2.2 $\mu$m is considerably smaller than at optical wavelengths and
the K-correction is relatively insensitive to the star formation history of 
the galaxies (e.g. Madau, Pozzetti \& Dickinson 1998).

Knowledge of the X-ray luminosity of a cluster is vitally 
important in the study of BCG evolution (Edge 1991).  X-ray parameters provide 
objective information about the cluster environment since the strength of 
X-ray emission is directly related to the depth of the cluster gravitational potential 
well and therefore the cluster mass. In both Collins \& Mann (1998) and Burke, Collins \& Mann (2000 -- hereafter BCM)
 it was established that, in general, BCGs in clusters with an X-ray luminosity 
$L_{\rm{X}}\geq 2.3\times10^{44}$ erg s$^{-1}$ (in the EMSS passband $0.3-3.5$ keV) 
have mean absolute magnitudes $\simeq0.5$ mag brighter with a dispersion half as 
large ($\simeq0.24$) as those in less luminous clusters.  Despite these efforts,
 it is unclear whether this effect is evolutionary as there are few BCGs with 
redshifts below $z\simeq0.1$ in these studies. For example, in 
Arag\'{o}n-Salamanca et al. (1998) there are only 3 BCGs below $z=0.1$ (2 
in high-$L_{\rm{X}}$ clusters) and 13 in BCM (3 in high-$L_{\rm{X}}$ 
clusters). To ascertain whether the variation of BCG properties with cluster 
X-ray luminosity seen at high redshift is a genuine evolutionary
effect we significantly extend the K-band Hubble diagram 
of BCM below $z\simeq0.1$ in this letter, using 
the 2 Micron All Sky Survey second incremental release extended source 
catalogue (hereafter 2MASS catalogue; Jarrett et al. 2000).

In Section 2 we give a brief outline of the 2MASS catalogue, describe our 
sample and then present the new Hubble diagram in Section 3.  In Section 4 we
 discuss the implications of our results for the evolution of BCGs and draw our
 conclusions. Throughout this letter we assume a cosmology consistent with the 
supernovae data (see Bahcall et al. 1999): $\Omega_{\rm{m}}=0.3, \Omega_{\rm{\Lambda}}=
0.7$ and H$_{\rm{0}}=100~h$ km s$^{-1}$ Mpc$^{-1}$, with $h=0.7$.  This affords the 
chance to present earlier results for the Hubble diagram, which assumed an 
Einstein--de Sitter universe and $h=0.5$, in the framework 
of the most recent estimates of the cosmological parameters.

\section{Data}

The 2MASS catalogue (Jarrett et al. 2000) covers $\sim$40 per cent of the sky 
and contains J, H and K$_{\rm{s}}$-band images and photometry for $\sim$585,000 
galaxies. The extended source detection limit (10$\sigma$) at K$_{\rm{s}}$ is 13.1 
mag and together with the spatial resolution of 2 arcsec means that 2MASS is 
well suited for detecting luminous galaxies, such as BCGs, out to
$z\sim0.1$. The typical astrometric accuracy of sources in 2MASS is
$0.5$ arcsec rms.

Our input cluster catalogue is that of Lynam (1999). This consists of a sample 
of 150 rich Abell clusters with measured redshifts $z\leq0.1$ and
$|b|\geq 15^{\circ}$, which have 
extended X-ray emission above a flux limit of $3\times10^{-12}$ erg s$^{-1}$ 
cm$^{-2}$ in the {\it ROSAT} hard-band ($0.5-2.0$ keV). The X-ray data comes from the
 second reduction of the {\it ROSAT} All Sky Survey, paired with co-ordinates from 
Abell, Corwin \& Olowin (1989). X-ray flux measurements are essentially the 
same as that of the NORAS and REFLEX cluster surveys (B\"{o}hringer et al. 
2000). The Lynam sample 
was constructed specifically to study streaming flows of 
galaxy clusters (Lynam et al. 2002, in preparation) and contains clusters with 
a wide range in X-ray luminosity selected over a large fraction of the entire 
sky, thus fulfilling our requirements. The BCGs were identified by the 
positional coincidence between the centroid of the X-ray cluster
emission and the first ranked galaxy, which in general show very good
agreement being coincident to $\leq 30$ arcsec (Lazzati \& Chincarini 1998).

The matching with the 2MASS catalogue was done by searching around the BCG 
positions given by Lynam (1999) out to a maximum radius of 60 arcsec.
A total of 79 BCGs were common to both catalogues, this constitutes only 50 
per cent of the Lynam (1999) catalogue due to the partial sky coverage of the 
2MASS second incremental release.  Only 19 of these had separations $>3$ 
arcsec, and all had separations less than $20$ arcsec. These 19 BCGs were 
checked carefully to ensure that the correct object had been identified.  In 
3 other cases the 2MASS algorithm had missed the BCG but found another source 
nearby, possibly due to source confusion. As these BCGs had no 
reliable K magnitude it was not possible to include them in the analysis.

The major strategy behind the photometric reductions of the 2MASS data was to
 be as consistent as possible with the procedures of BCM for the 76 BCGs 
discussed in that paper. Out of the wide range of photometry values available 
in the 2MASS catalogue we chose to use those measured in circular apertures as 
they are more robust than those in elliptical apertures (e.g. Pahre 1999, 
Jarrett et al. 2000). The catalogue contains magnitudes within 11 circular 
apertures with radii between 5 and 70 arcsec.  A Hermite polynomial fit to the 
aperture photometry of the 2MASS data was performed to estimate the apparent 
magnitude at the metric radius of $r_{\rm{m}}=12.5h^{-1}$ kpc used by BCM.  The internal photometric error in the 2MASS magnitudes is $\sim\pm0.1$ mag.  The 
data was then corrected for Galactic absorption using the maps of Schlegel, 
Finkbeiner, \& Davis (1998) -- the correction is small, typically $\sim0.02$ 
mag. The colour correction between the 2MASS K$_{\rm{s}}$-band and the UKIRT 
K-band, used by BCM, is typically $0.006\pm0.01$ mag (Carpenter 2001), and so 
was neglected.  

One difference between the datasets is that in the BCM sample no attempt was 
made to remove flux due to other cluster galaxies falling within the aperture, 
although flux due to contaminating stars and obvious non-cluster galaxies was 
excluded.  In contrast, the 2MASS magnitudes are automatically corrected for 
the presence of stars and other galaxies within each aperture. We examined the 
difference that this makes by measuring the magnitudes of 10 2MASS galaxies in 
circular apertures directly from the 2MASS `postage stamp' images. The mean 
difference in the photometry between the uncorrected postage-stamp images and 
the 2MASS magnitudes is only $0.03\pm0.01$ mag. Finally, as an external 
check on the consistency of the 2MASS photometry, we obtained K-band 
magnitudes for 12 BCGs using QUIRC on the University of Hawaii 2.2m telescope. 
 The 12 BCGs cover the range of redshift and cluster X-ray luminosity values 
in the Lynam (1999) sample.  The mean difference (2MASS-UH) is only 
$0.028\pm0.042$ mag.  We also compared the magnitudes of the 6 BCGs below 
$z=0.1$ in BCM which are in common with 2MASS.  The mean difference 
(2MASS-BCM) is $0.04\pm0.09$. Therefore we conclude that there are no 
significant offsets between the respective datasets.

All X-ray luminosities were corrected to the new cosmology, denoted by 
$\Lambda$, using the form:

\begin{equation}
L_{\rm{\Lambda}}=L_{\rm{EdS}}\left( {{d_{\rm{L(\Lambda)}}\over 
{d_{\rm{L(EdS)}}}}} \times \frac{apcorr_{(\Lambda)}}{apcorr_{\rm (EdS)}}  \right)^2,
\end{equation}

where EdS denotes an Einstein--de Sitter cosmology with $h=0.5$, $d_{\rm{L}}$ 
is the luminosity distance to the cluster and $appcor$ is the value of
the aperture correction. The BCM data incorporates X-ray
catalogues, principally EMSS (Gioia \& Luppino 1994) and the SHARC
surveys (Burke et al. 1997; Romer et al. 2000), which extrapolate
their measured X-ray flux to total flux assuming a King surface
brightness profile and a fixed core radius. The appropriate 
value of $apcorr$ was used for each survey (giving rise to luminosity
corrections $\simeq20$ per cent), neglecting the RASS-based 
measurements as these 
fluxes are $\geq 90\%$ of the total flux from the cluster (B\"{o}hringer et 
al. 2000, 2001) resulting in a negligible aperture correction. Finally, to 
maintain consistency with BCM, the X-ray luminosities given by Lynam (1999) 
were transformed from the RASS ($0.1-2.4$ keV) passband to the EMSS passband 
($0.3-3.5$ keV).  Assuming a 6 keV thermal bremsstrahlung spectrum (the 
choice of temperature has negligible effect on the correction factor): the 
passband corrections are of the form: $L_{\rm{X}}(0.3-3.5$ keV$)=1.08 L_{\rm{X}}
(0.1-2.4$ keV). At the mean redshift of the sample ($z=0.27$), the X-ray luminosity 
$2.3\times10^{44}$ erg s$^{-1}$ ($0.3-3.5$ keV), reported in BCM as marking a
transition in the properties of the constituent BCGs, transforms to
$\simeq1.9\times10^{44}$ erg s$^{-1}$ in the supernovae cosmology
adopted here. Hence we report the results below using this new X-ray
luminosity threshold.     


\section{Results}

\begin{figure*}
\epsfig{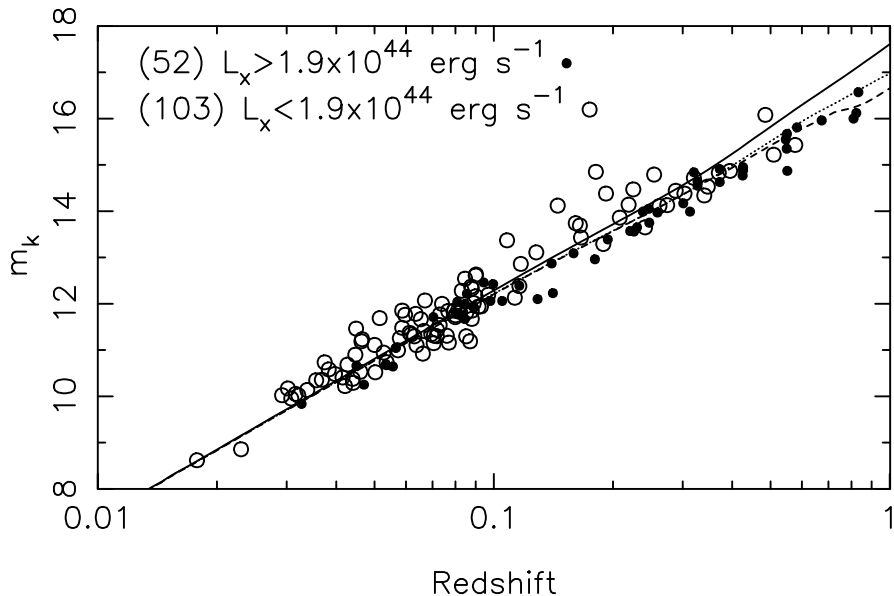}
\caption{The Hubble diagram. The filled points denote BCGs in clusters 
with $L_{\rm{X}}(0.3-3.5$ keV$)>1.9\times10^{44}$erg s$^{-1}$. The no-evolution 
prediction, assuming a 10 Gyr old stellar population, is shown by the solid 
line, the $z_{\rm{f}}=2$ model by the dashed line and the $z_{\rm{f}}=5$ model 
by the dotted line. All the models were normalised in the same way using the 
18 BCGs below $z=0.1$ with $L_{\rm{X}}\geq1.9\times10^{44}$ erg s$^{-1}$ which 
gives $M_{\rm{k}}=-25.78\pm0.05$.}
\label{cross}
\end{figure*}

The K-band Hubble diagram for 155 BCGs is shown in Figure 1.  The lines show 
model predictions calculated using the GISSEL96 code (Bruzual \& Charlot 1993),
 for a solar metallicity stellar population with a Salpeter initial mass 
function. The solid line indicates a no-evolution model for a 10 Gyr old 
stellar population, while the other lines are for stellar populations which 
form in an instantaneous burst of star formation at a single epoch, $z_{\rm{f}}=2$ and 
$z_{\rm{f}}=5$, and then evolve passively.  

In Figure 2 we present the absolute magnitude -- X-ray luminosity relation for 
the high and low redshift samples. The $M_K$ values are calculated
using the K-correction from the $z_{\rm{f}}=2$ model. The figure illustrates 
clearly how the $M_K$ values for BCGs at $z\geq0.1$ depend on the
X-ray luminosity of the host cluster and contrasts this behaviour with the 
$M_K$ distribution at $z\leq0.1$. This 
difference is quantified in Table 1 which gives the mean absolute magnitudes 
as a function of X-ray luminosity and redshift.  To test the significance of 
this difference we carried out a non-parametric two-tailed Kolmogorov--Smirnov 
(KS) test on the absolute magnitudes of the BCGs in clusters above and below 
$z=0.1$.  Above $z=0.1$ the test gives a probability of only 0.013 per cent 
that BCGs in high and low $L_{\rm{X}}$ clusters are drawn from the same 
parent population, whereas the corresponding probability for BCGs below
 $z=0.1$ rises to $36$ per cent.

\begin{figure}
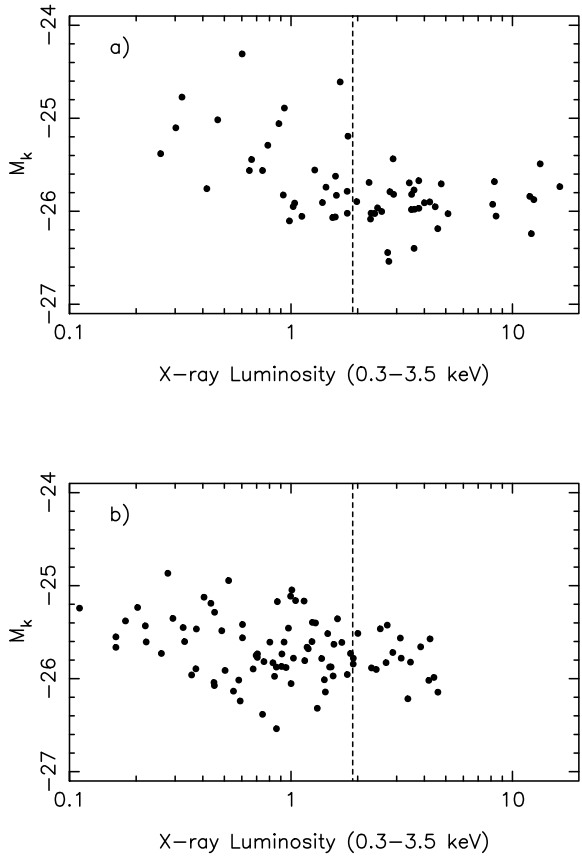

\epsfig{file=mkvslx_highz.eps,angle=-90,width=9.0cm}
\epsfig{file=mkvslx_lowz.eps,angle=-90,width=9.0cm}
\caption{The absolute magnitude - X-ray luminosity relation for a) those BCGs 
with redshift $z\geq0.1$ and b) those with redshift $z\leq0.1$. The
dotted line represents the $L_{\rm{X}}=1.9\times10^{44}$ erg s$^{-1}$ 
cluster luminosity boundary.}
\end{figure}
\vspace{0cm}

\begin{table}
\centering
\caption{Statistical properties of the absolute magnitude as a function of 
X-ray luminosity and redshift.}
\begin{tabular}{lllc}
\multicolumn{1}{c}{z} & \multicolumn{2}{c}{$L_{\rm{X}} (0.3-3.5$ keV)
$\times10^{44}$erg s$^{-1}$ (N, $1 \sigma$)}\\
& \multicolumn{1}{c}{$<1.9$} & \multicolumn{1}{c}{$>1.9$}\\ 
\hline
$<0.1$&-25.67$\pm0.04$ (74, 0.34)&-25.78$\pm0.05$ (18, 0.21)\\ 
$>0.1$&-25.51$\pm0.09$ (29, 0.47)&-25.93$\pm0.04$ (34, 0.24)\\ 
\hline
\end{tabular} 
\end{table}

\section{Discussion and Conclusions}

We consider the amount of merging that BCGs may have undergone to further 
examine the difference in BCG properties with host cluster environment. The 
parametric form introduced by Arag\'{o}n-Salamanca et al. (1998) and used by BCM: 
$M(z)=M(0)\times(1+z)^{\gamma}$ was used to estimate the growth in the stellar 
mass content of the BCGs.  Figure 3 shows the residuals about a $z_{\rm{f}}=2$ model
 with the lines corresponding to growth factors of 0, 2, 4, 16 since
$z=1$.  
It is clear from Figure 3a that those BCGs in more X-ray luminous clusters have
 not experienced any significant stellar mass growth since $z=1$, whereas 
Figure 3b indicates that the BCGs in low X-ray luminosity clusters show a 
wide range of mass evolution since redshift $z=1$.  Very similar results are 
found if the formation redshift is increased to $z_{\rm{f}}=5$. This
suggests that BCGs have different evolutionary histories depending on
the X-ray properties of their cluster hosts.

\begin{figure}
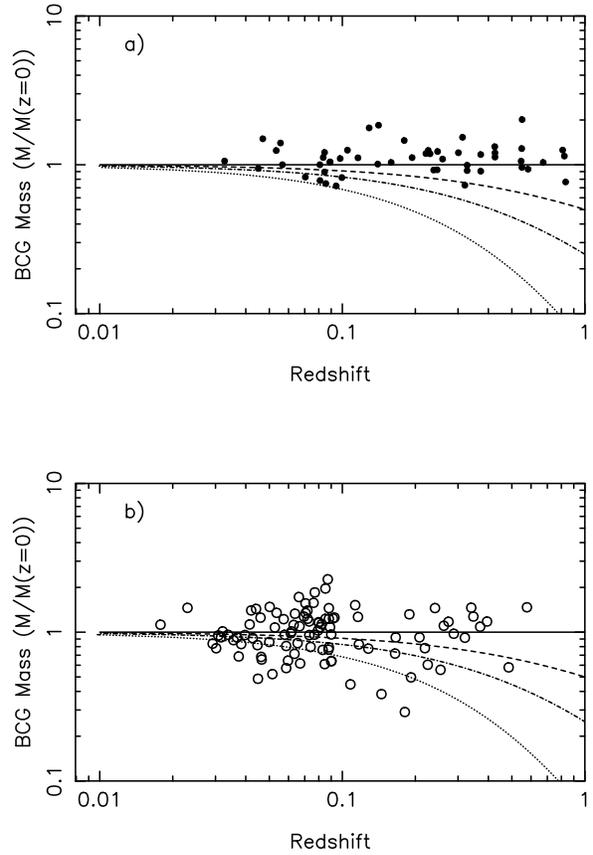

\epsfig{file=growth.eps,angle=-90,width=9.0cm}
\epsfig{file=growth2.eps,angle=-90,width=9.0cm}
\caption{Inferred mass of BCGs from residuals about the $z_{\rm{f}}=2$ model for 
BCGs in a) high-$L_{\rm{X}}$ clusters and b) in low-$L_{\rm{X}}$ clusters.  The model 
plots are for $M(z)=M(0)(1+z)^{\gamma}$, with $\gamma=0,-1,-2, -4$, corresponding 
to growth factors of 0, 2, 4, 16 since $z=1$. The models are normalised
independently for each figure using the absolute magnitudes below
$z=0.1$.}
\label{growth}
\end{figure}
\vspace{0cm}

Our results on the merging 
history of BCGs have been qualitatively confirmed recently by Nelson et al. 
(2001) using the I-band Hubble diagram for BCGs selected from the Las Campanas 
Distant Cluster Survey in the range $0.3\leq z \leq 0.9$.  In this sample only 
BCGs in low mass clusters must have accreted significantly since $z\simeq1$. 
Furthermore, the V-I colour evolution seen by Nelson et al. (2001) is 
consistent with the accretion in low-$L_{\rm{X}}$ BCGs consisting of old stellar 
populations. From a theoretical perspective, N-body simulations 
naturally produce massive, central dominant cluster galaxies through merging 
(Dubinski 1998, Athanassoula, Garijo \& Garc\'{i}a G\'{o}mez 2001), although relatively little theoretical work has
focussed on predicting the merging history of BCGs as a function of
environment: the semi-analytical models presented by Arag\'{on}-Salamanca 
et al. (1998), predict that BCGs in the most massive clusters have grown by a 
factor $4-5$ since $z=1$, which is clearly inconsistent with the results in 
Figure 3a. More recently, Gottl\"{o}ber, Klypin \& Kravtsov (2001) argue 
that at $z\sim1$ the merger rates in groups are significantly higher than in
richer systems, resulting in a more diverse range of evolutionary
properties of their constituent galaxies. 
 
We conclude that those BCGs in high redshift, high luminosity clusters are 
brighter and more uniform than those in their low luminosity counterparts and 
that this separation is significantly weaker at low redshift. This is
a direct indication of how a single homogeneous population
of galaxies evolves hierarchically. Both this work and 
future studies of the near-infrared surface brightness profiles of BCGs along 
with velocity dispersion measurements should help constrain the evolutionary 
parameters and stimulate more theoretical predictions.

\section*{Acknowledgments}
SB acknowledges PPARC for a Postgraduate Studentship.  DJB acknowledges the support of SAO contract SV4-64008 and NASA contract NAS8-39073(CXC).  RGM acknowledges support from PPARC.  This publication makes use of data products from the Two Micron All Sky Survey (2MASS) which is a joint project of the University of Massachusetts and the Infrared Processing and Analysis Center/California Institute of Technology, funded by the National Aeronautics and Space Administration and the National Science Foundation.  We acknowledge the use of the UH 2.2m telescope at the Mauna Kea Observatory, Institute for Astronomy, University of Hawaii.

\label{lastpage}

\bsp

\end{document}